\documentclass[conference]{IEEEtran}
\IEEEoverridecommandlockouts

\usepackage{cite}
\usepackage{amsmath,amssymb,amsfonts}
\usepackage{algorithm}
\usepackage{algorithmic}
\usepackage{graphicx}
\usepackage{textcomp}
\usepackage{xcolor}
\usepackage{booktabs}
\usepackage{url}
\usepackage[skip=15pt]{caption}
\usepackage{subcaption}
\usepackage{hyphenat}
\usepackage{fancyhdr}
\usepackage[top=0.70in, bottom=1.05in, left=0.673in, right=0.673in]{geometry} 
\def\BibTeX{{\rm B\kern-.05em{\sc i\kern-.025em b}\kern-.08em
    T\kern-.1667em\lower.7ex\hbox{E}\kern-.125emX}}
\usepackage{fancyhdr}
\begin{document}
\bstctlcite{IEEEexample:BSTcontrol}

\title{
Quantifying the Energy-Saving and QoS Trade-Off in Traffic Offloading for Real 4G/5G Scenarios\\
\thanks{This work is supported by the Grant TRAINER-6G (PID2023-146748OB-I00) funded by MCIN/AEI/10.13039/501100011033 and by ERDF/EU, and by the Smart Networks and Services Joint Undertaking (SNS JU) under the European Union’s Horizon Europe research and innovation programme under Grant Agreement No 101097083, BeGREEN project. Views expressed are however those of the author(s) only and do not necessarily reflect those of the European Union or SNS-JU. Neither the European Union nor the granting authority can be held responsible for them. }
}
\author{
    \IEEEauthorblockN{David Reiss\IEEEauthorrefmark{1}\IEEEauthorrefmark{2}, Miguel Catalan-Cid\IEEEauthorrefmark{1}, Daniel Camps-Mur\IEEEauthorrefmark{1}, Oriol Sallent\IEEEauthorrefmark{2}}
    \IEEEauthorblockA{\IEEEauthorrefmark{1}i2CAT Foundation, Spain. \{david.reiss, miguel.catalan, daniel.camps\}@i2cat.net}
    \IEEEauthorblockA{\IEEEauthorrefmark{2}UPC, Spain. \{david.reiss, jose.oriol.sallent\}@upc.edu}
}

\maketitle
\thispagestyle{fancy} 

\begin{abstract}
Despite the potential for higher energy efficiency in 5G networks, current 5G Non-Standalone (NSA) deployments often operate suboptimally due to low utilization of 4G and 5G carriers during extended periods. Since base stations are the primary contributors to network energy consumption, implementing cell on/off switching and traffic offloading strategies is crucial for enhancing energy efficiency in current deployments. This paper investigates energy-saving opportunities based on these strategies in a real 5G NSA deployment, utilizing a dataset provided by a European Mobile Network Operator. Using Key Performance Indicators from the dataset, we propose a data-driven framework to evaluate the energy-saving and QoS trade-off when selectively deactivating underutilized 5G cells and offloading their traffic to 4G cells with enough resources within the same sector and site. Our results demonstrate network-wide cell switch-off opportunities ranging from 17\% to 79\%, while ensuring data rates between 25 Mbps and 5 Mbps, respectively.
\end{abstract}


\section{Introduction}

{
Developing transversal and sustainable solutions across different industries is one of the most significant challenges in combating climate change and its consequences. In this context, mobile communications 
represent around 0.3\% of the global GHG emissions, equivalent to 140 million tonnes \cite{gsma}.}{ Although energy per bit has been improved compared to LTE, the active Massive Multiple Input Multiple Output (MIMO) antennas used in 5G deployments lead to excessive energy consumption during low demand periods. 
Indeed, the Radio Access Network (RAN) remains the most energy-demanding part of 5G networks, accounting for 75\% of its energy consumption \cite{gsma}. 

The 3GPP study on energy savings for New Radio (NR), provides a comprehensive overview of various approaches to address this challenge \cite{3GPP_esavings}
. In the RAN domain, the on/off switching approaches are often used to
exploit low demand conditions. These strategies can be applied at different  levels, e.g., deactivating RF channels, sectors, carriers, or entire sites can be achieved by switching off components such as the Power Amplifier (PA), the Radio Unit (RU), the node, or the entire Base Station (BS), respectively \cite{esaving_techonologies}. For instance, carrier and cell on/off switching is one of the main energy saving use cases considered by the O-RAN Alliance, as detailed in \cite{oran-esaving-lates}.

Two main challenges emerge when applying switch-off strategies in operational networks. The first is the availability of realistic and extensive datasets. Such datasets enable the characterization of 5G adoption \cite{real-adoption} and support the application of traffic forecasting methods \cite{traffic-forecast}, which are key to effectively evaluate the benefits and impacts of energy savings strategies. For instance, a study based on a real-world dataset from China in \cite{real-energy-efficiency} revealed that, due to the high baseline consumption of 5G radio equipment, 5G base stations become less efficient than 4G ones under low demand conditions. To address this, the authors explored channel and carrier shutdown jointly with deep sleep modes to enhance energy efficiency in 5G network. In this paper, we adopt a similar approach while addressing the second critical challenge: balancing energy savings with the Quality of Service (QoS).



In \cite{es-qos-commag}, the authors explore the trade-off between  CO$_2$ emissions and the QoE (Quality of Experience) of multimedia services, demonstrating that reducing the carbon footprint comes at the cost of sacrificing certain degree of service quality. As emphasized in \cite{sustainability_6G}, strategies to address and manage this trade-off will play a pivotal role in 6G standardization.
Authors in \cite{ml-intelligent-ps}, propose a cell on/off energy-saving strategy evaluated using an open-source dataset from an Italian operator. The study highlights that considering the operator's Service Level Agreement (SLA) outage significantly constrains energy savings. However, the SLA outage definition only considers erroneous switch-off decisions that may lead to saturation of other cells, without assessing the impact on specific performance indicators.  


In this context, the main contribution of this paper is to provide a framework to evaluate the energy-saving and QoS trade-off when selectively deactivating underutilized 5G cells and offloading their traffic to 4G cells. The study, which is based on the analysis of a dataset from a commercial european Mobile Network Operator (MNO), focuses on three aspects. First, we study the upper bound of achievable energy savings in the deployment characterized by the dataset. Second, we assess the impact of imposing certain QoS limitations during the offloading procedure, defined as minimum data rate levels to be ensured for offloaded UEs. Finally, we investigate the influence of parameters governing cell switch on/off operations on the energy-saving and QoS trade-off. Results highlight the importance of considering site-specific characteristics and performance indicators, discouraging the use of global fixed strategies for effective optimization.


The paper is structured as follows. Next section provides the dataset description, highlighting the most relevant  Key Performance Indicators (KPIs). Section \ref{sec:qos} describes the framework used to analyze the trade-off between energy savings and QoS. Section \ref{sec:eval} provides the results of a network-wide evaluation. Finally, we present the conclusions and the future work.  

\section{Dataset Description and Analysis}

{The performed study is based on a dataset provided by a European MNO, which includes a comprehensive list of KPIs from the RAN of a real cellular network. The deployment covers an extensive area of a large European city and its surroundings, including both urban and suburban environments. The analyzed region is divided into two areas, as depicted in Figure \ref{fig:deployment}. The smaller area 
corresponds to a dense urban area within the city, covering an approximate extension of 14 $km^2$. The larger area includes various cities, leading to a mix of urban and suburban scenarios, covering approximately 100 $km^2$.}

\begin{figure}[t]
    \centering
    \vspace{0.04in}
    \includegraphics[width=0.55\linewidth]{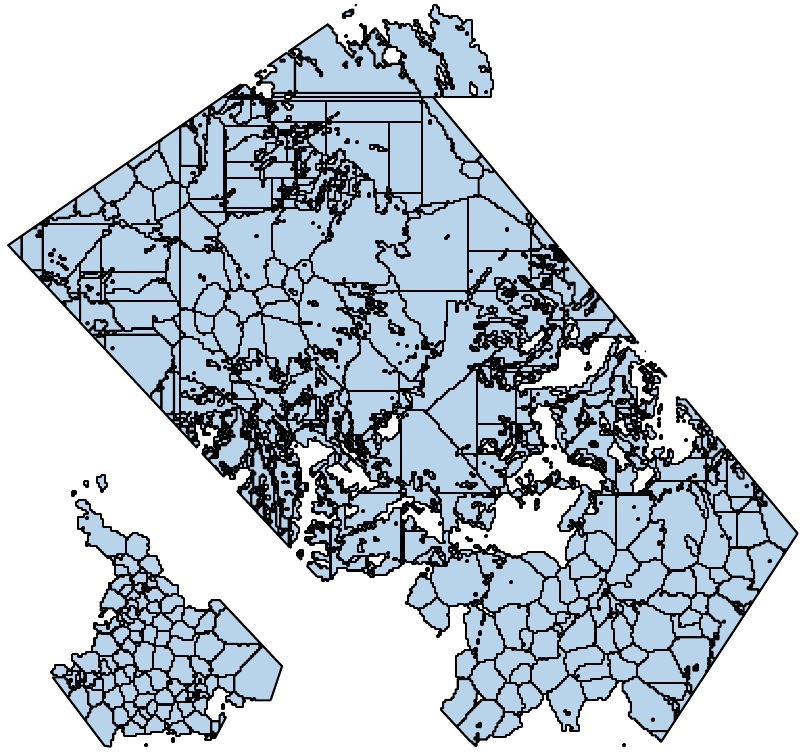}
    \caption{Region covered by the dataset, illustrating the coverage area of the 3500 MHz 5G cells.}
    \label{fig:deployment}
\end{figure}

{The dataset 
spans two full consecutive months, plus additional non-consecutive weeks,
and the reported KPIs are the average values over the preceding 15 minutes. The number of KPIs, sites, carriers, and cells varies depending on the Radio Access Technology (RAT). For 4G, the dataset includes 312 sites and 3427 cells (with up to 5 carriers per site), resulting in a total of 1314 KPIs. In contrast, 5G contributes 220 sites and 1271 cells (with up to 3 carriers per site), accounting for 679 KPIs.
Throughout the paper, we will use the term \textit{node} to refer to the group of cells or sectors associated with a specific carrier at a site, which is usually three (one per sector). 

{The traffic offloading strategy and evaluation proposed in this paper rely on a specific subset of KPIs, 
which are detailed in Table \ref{tab:KPI-description}.} 
{The energy-related KPIs allow us to estimate the baseline consumption of the 5G nodes and the potential energy savings. The remaining KPIs are utilized to implement the traffic offloading strategy and to characterize its impact on QoS. The study focuses on the downlink (DL), as it is the most demanding traffic direction. It assumes that whenever the DL load can be offloaded, offloading the uplink (UL) load will also be possible\footnote{The analysis of the dataset revealed that the aggregated UL load demand of 4G and 5G remained always below 50\% of the available 4G resources.}.}

\smallskip
\begin{table}[h]
\captionsetup{skip=1pt}
\caption{Relevant KPI Description.}
\begin{center}
\begin{tabular}{|p{1.5 cm}|p{4.3 cm}|p{1.3 cm}|}
\hline
\textbf{Name} & \textbf{Description} & \textbf{Technology}  \\
\hline
{Consumed \nohyphens{Energy}} & {Energy consumption of the nodes (Wh)}  & 5G \\
\hline
{Daily Consumption} & {Daily aggregated energy consumption (kWh)} & 5G, 4G  \\
\hline
{Average DL Load} & {Average downlink (DL) load of the past 15-minute interval (\%)}   & 5G, 4G\\
\hline
{Average DL Throughput per UE} & {Average throughput per user equipment (UE) in the past 15-minute interval (Mbps)} & 5G, 4G \\
\hline
{{Average Connected UEs}} & {Average number of 5G connected UEs, without specifying their states} & 5G\\
\hline
{Cell Name} & {Name of the cell, specifying the RAT, carrier, and sector of a given cell}  & 5G, 4G \\
\hline
\end{tabular}
\label{tab:KPI-description}
\end{center}
\end{table}

\begin{table}[h]
\captionsetup{skip=1pt}
\caption{4G and 5G radio features.}
\begin{center}
\begin{tabular}{|c|c|c|c|c|}
\hline
\textbf{Carrier} & \textbf{BW} & \textbf{SCS} & \textbf{UL/DL PRBs} & \textbf{Duplexing} \\
\hline
{4G 700 MHz} & 10  & 15 & 50/50 & FDD \\
\hline
{4G 800 MHz} & 10  & 15 & 50/50 & FDD \\
\hline
{4G 1800 MHz} & 20  & 15 & 100/100  & FDD \\
\hline
{4G 2100 MHz} & 10  & 15 & 50/50 & FDD \\
\hline
{4G 2600 MHz} & 20  & 15 & 100/100 & FDD \\
\hline
{5G 3500 MHz} & 100  & 30 & 68/205 & TDD \\
\hline
\end{tabular}
\label{tab:rfeature}
\end{center}
\end{table}

{Due to the high correlation observed between traffic demand from different weeks, the study presented in the following sections is simplified and based on data from just one specific week. The analysis revealed that the behavior of all examined KPIs was consistent, following a cyclic pattern strongly correlated with day-night cycles. Using a complete week captures differences between weekdays and weekend days, particularly Sunday, when network demand typically decreases.}

\subsection{Energy consumption and utilization analysis}
On/off cell switching strategies are commonly based on deactivating the capacity layer \cite{capacity_layer}. In the targeted scenario, the capacity layer corresponds to the 3500 MHz 5G carrier, which provides the highest bandwidth, as shown in Table \ref{tab:rfeature}. Notably, the other 5G carriers (700 MHz and 2100 MHz) are deployed using Dynamic Shared Spectrum (DSS) and share radio equipment with 4G, so they cannot be switched off independently. Additionally, their utilization levels were marginal when compared to the 3500 MHz carrier.

{Table \ref{tab:rfeature} provides a summary of the most relevant radio features of 4G and 5G cells, including the number of Physical Resource Blocks (PRBs) determined by the bandwidth (BW) and the Sub Carrier Spacing (SCS).
For 5G, as indicated by the operator, we considered a 25\%/75\% UL/DL rate, which corresponds to a TDD pattern of DDDSU, with a special slot, $S$, configured with a ratio of 10 Downlink symbols, a 2-symbol Guard Period, and 2 Uplink symbols \cite{gsmatdd}. 


Figure \ref{fig:energy} illustrates the daily energy consumption of the nodes, comparing the aggregate consumption of all 4G carriers with that of the 5G 3500 MHz carrier. As shown in the figure, the average energy consumption in both cases is comparable, although the deviation in 4G is significantly higher. This variation is mainly due to the different number of active carriers across sites, driven by cell planning strategies and occasional 4G sleep cell policies\footnote{We observed occasional deactivation of some 4G carriers, predominantly during nighttime hours, decreasing the availabilty of 4G resources in some sites.}. Note that 5G typically offers higher bandwidth than the aggregate bandwidth of the 4G carriers, highlighting the potential higher energy efficiency of 5G technology \cite{capacity_layer}.



\begin{figure}[t]
    \centering
    \begin{subfigure}[t]{0.48\columnwidth} 
        \centering
    \centering
    \includegraphics[width=0.95\linewidth]{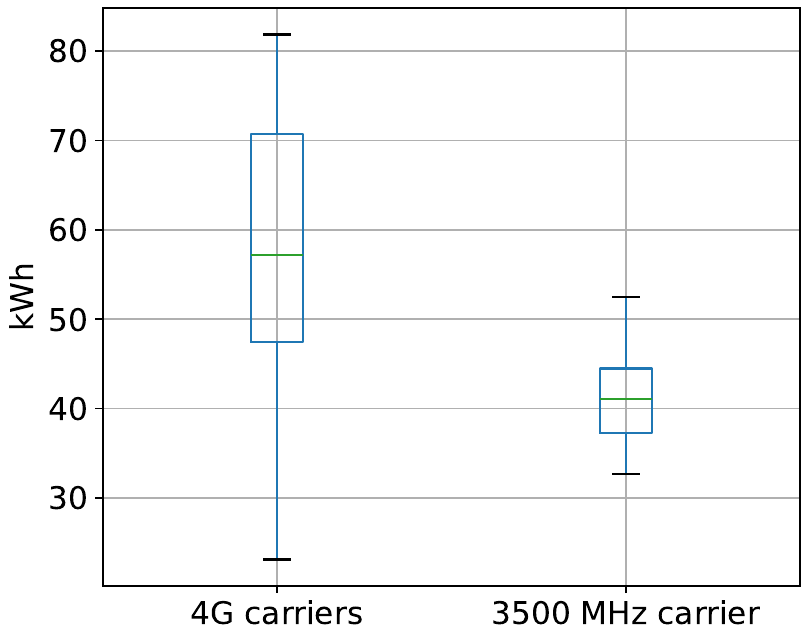}
    \caption{Energy consumption}
    \label{fig:energy}
    \end{subfigure}%
    \hspace{+1mm}
    \begin{subfigure}[t]{0.48\linewidth} 
        \centering
    \includegraphics[width=0.95\linewidth]{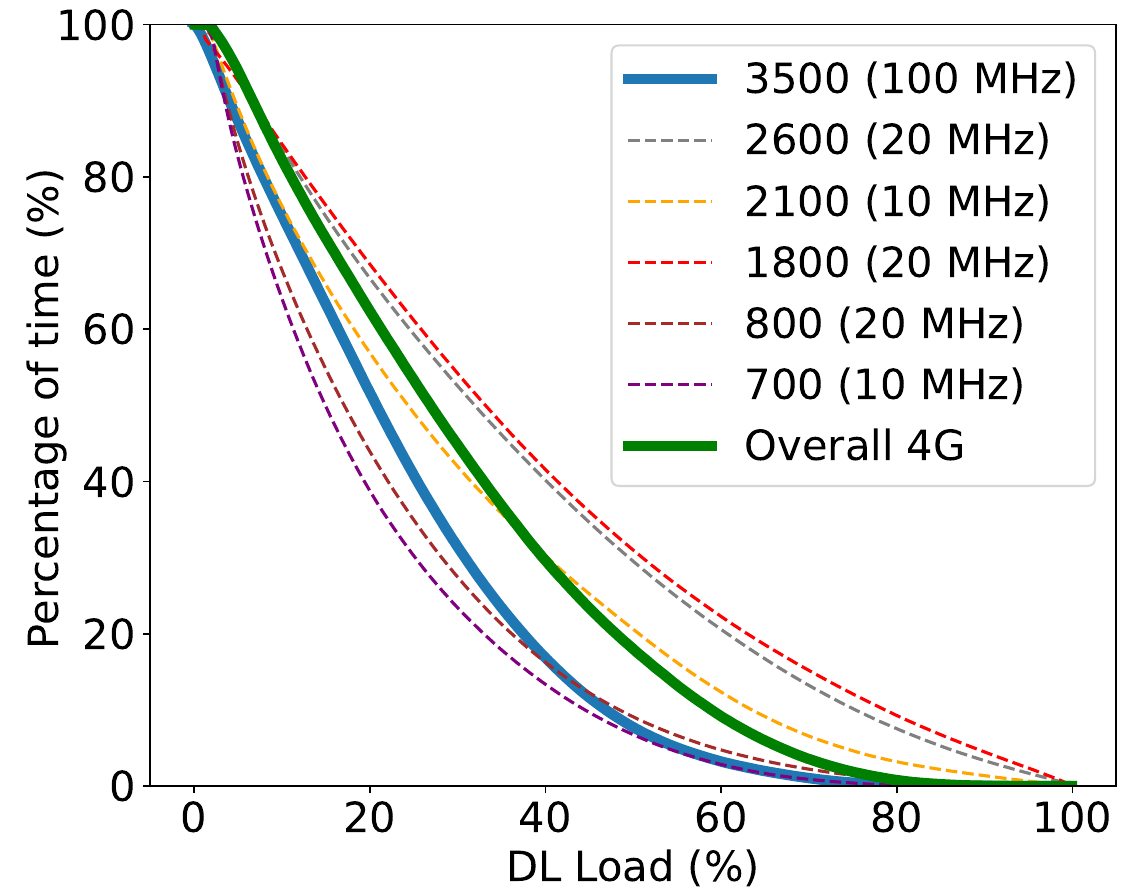}
    \caption{Utilization}
    \label{fig:utilization}
    \end{subfigure}
    \caption{Comparison of 4G and 5G carriers.}
    \label{fig:5gvs4g}
\end{figure}



{Figure \ref{fig:utilization} depicts the  Complementary Cumulative Distribution Function (CCDF) of the utilization of 4G and 5G carriers, illustrating the percentage of time above each load level. The figure shows that the 5G carrier's utilization is significantly lower than that of the most heavily used 4G carriers and is comparable to the least utilized ones. Furthermore, it falls below the overall 4G utilization, which is calculated as the total aggregated 4G resource demand divided by the total available 4G resources. This indicates that, despite its potential to be more energy-efficient than 4G cells, the current 5G deployment analyzed in this dataset is underutilized, resulting in poor energy efficiency. Therefore, selectively deactivating 5G carriers and offloading their traffic to 4G cells would not only decrease the total energy consumption of the RAN but also substantially enhance its energy efficiency.


\subsection{Estimation of the saved energy}\label{sec:energy-baseline}

{The energy savings estimation depends on both 4G and 5G cell's energy consumption. Since the dataset does not include  energy data of the 4G nodes, we assume that the additional energy consumed by the 4G cells after the offloading (i.e., due to increased traffic demand) approximates the energy previously consumed by the 5G cells for serving the same traffic. Therefore we estimate the energy savings as the baseline consumption of the deactivated 5G cells. Although the previous assumption may cause a marginal error to appear in the estimation, note that the baseline consumption represents from 60\% to 100\% of the total energy consumption (depending on load conditions, see Figure \ref{fig:energy-load}). Hence, it represents the main contribution to the energy savings, specially during off-peak hours where most of switch-off cases will happen. } 

{Figure \ref{fig:energy-load} shows the correlation between load and energy KPIs. During low demand periods, where energy contribution due to traffic demand is negligible, the reported energy approximates the baseline consumption of the 5G nodes. Particularly, for the analyzed node it is found approximately at 400 Wh every 15 minutes, or 1600 Wh per hour. Then, the baseline consumption per cell is equal to 133 Wh every 15 minutes, or 532 Wh per hour, i.e., one third of the node's baseline consumption. }



{In the previous estimation we do not account for the energy split between the Radio Unit (RU) and the Base-Band Unit (BBU). According to \cite{ntt-greennets}, in 5G commercial networks, the RU typically accounts for 88\% of the energy consumption on average and up to 78\% under maximum cell load conditions. Considering these aspects could slightly modify the energy saved estimation.
In any case, results in Section \ref{sec:eval} will mainly focus on the percentage of time that 5G cells could be deactivated, offering a broader and more generalizable perspective on the energy-saving benefits and trade-offs associated with the proposed traffic offloading strategy.

\begin{figure}[t]
    \centering
    \includegraphics[width=1\linewidth]{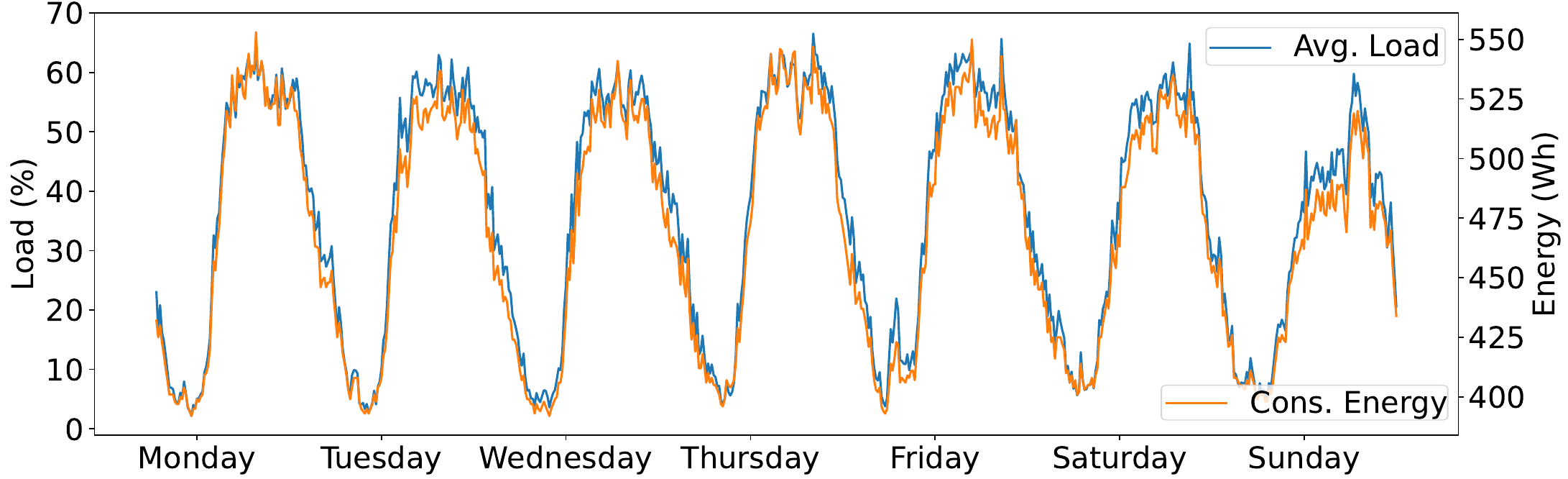}    \caption{Average load (blue) and energy consumption (orange) of one 5G 3500 MHz node.}
    \label{fig:energy-load}
\end{figure}


\section{Framework to Analyze Energy-QoS Trade-off during Traffic Offloading}\label{sec:qos}

In this section, we describe the framework designed to analyze the trade-off between energy savings and QoS when offloading traffic between 5G and 4G cells. First, we describe the methodology used to trigger the activation and deactivation of 5G cells, which determines the upper bound of energy-saving opportunities when QoS penalties are not taken into account. Next, we explain how the impact on QoS can be evaluated using the dataset KPIs and integrated into the offloading decision.

\subsection{Impact of traffic offloading on energy savings}

The selective deactivation of 5G cells requires offloading their active traffic to 4G cells. Specifically, we assume a generic cell switch-off approach that considers offloading between the 4G and 5G cells within the same sector and site. This ensures that offloaded UEs will maintain similar or higher signal levels according to the larger coverage area of 4G carriers. It should be noted that the dataset does not include information on the location of UEs, which limits the application of load-balancing strategies across different sectors. Implementing cross-sector strategies in deployments that support advanced features, such as traffic steering xApps in O-RAN architectures, could further enhance energy-saving opportunities by distributing traffic among sectors \cite{oran-esaving-lates}. 


To illustrate the designed traffic offloading strategy, Figure \ref{fig:demands} shows the load pattern of a representative urban cell with medium-high traffic demand. The blue trend represents the aggregated load across all 4G carriers, as defined in previous section, while the orange trend represents the 5G load. Both trends exhibit a strong correlation and follow the typical day-night traffic cycles. 
The green trend represents the total traffic demand on the 4G cells, including the offloaded traffic from the 5G carrier. This trend is calculated by converting 5G PRBs to 4G PRBs based on the difference in SCS configuration outlined in Table \ref{tab:rfeature}: i.e., the 5G slot duration is half the duration of the 4G slot. For example, when the 5G demand for a cell reports an average usage of 60\%, it corresponds to 123 PRBs per 5G slot (calculated as $60\% \times 205$). This translates to an equivalent 4G load of 246 PRBs per 4G slot ($123 \times 2$). As shown in the figure, this approach allows traffic offloading only when both the 4G and 5G loads are moderate to low, which typically occurs during night hours, depending on the number of active 4G carriers.

{Once the aggregation of 4G and offloaded 5G loads is calculated, the offloading decision becomes straightforward. If the aggregated value exceeds the defined threshold, $\gamma$,  (represented by the dashed line), the 5G cell remains active to handle the traffic demand. Otherwise, the 5G cell can be switched off to save energy. We refer to this strategy as the greedy switch-off strategy with parameter $\gamma$. Throughout this section, we have applied $\gamma = 100\%$ to explore the upper bound of energy savings, but other $\gamma$ values and their impact on QoS will be analyzed in Section \ref{sec:eval}.  

Following this strategy the example cell shown in Figure \ref{fig:demands} achieved an upper bound switch-off time equal to 53\% of the week. By applying the calculations introduced in \ref{sec:energy-baseline}, this corresponds to an energy saving of 47.4 kWh. Since this cell is located in a high-demand area, the upper bound for energy savings across the entire region is expected to be even higher, as detailed in Section \ref{sec:eval}.

\begin{figure}[t]
    \centering
    \vspace{0.04in}\includegraphics[width=1\linewidth]{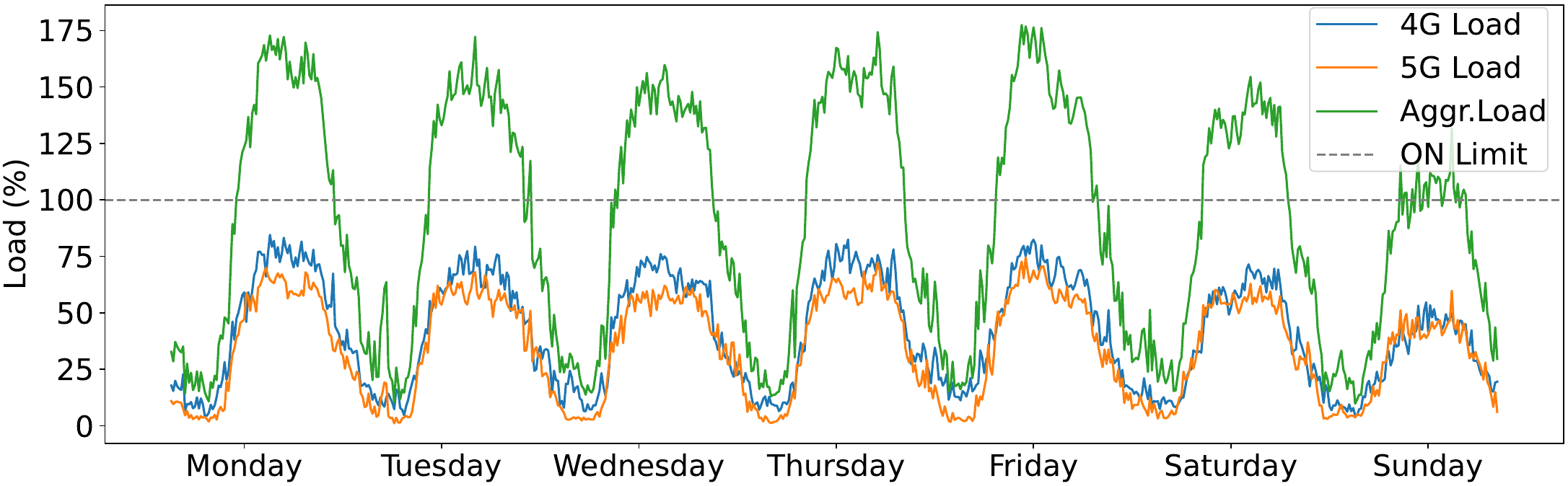}
    \caption{Load of a selected cell during the analyzed week in percentage of total PRBs: Aggregated 4G (blue), 5G (orange), and aggregated plus offloaded demand in 4G  (green). }
    \label{fig:demands}
\end{figure}

\subsection{Impact of traffic offloading on QoS}\label{sec:qos-meto}

{Notice that the greedy switch-off strategy  will result in a QoS penalty for the affected UEs. Although it ensures the full allocation of offloaded 5G PRBs, a reduction in uplink and downlink transmission rates due to differing capabilities of 4G and 5G cells will be experienced. This will impact, for instance, the time needed to download a file, and it could degrade the QoE of services which require specific data rates, such as video streaming. To this end, we analyzed the \textit{average downlink throughput per UE} KPI, which represents the experienced downlink transmission rate. Figure \ref{fig:scatter-thr} shows the scatter plot of the average load versus the average throughput per UE of a 4G 2600 MHz cell and a 5G 3500 MHz cell. 

\begin{figure}[t]
    \centering
    \begin{subfigure}[t]{0.48\columnwidth} 
        \centering
         \includegraphics[width=0.95\linewidth]{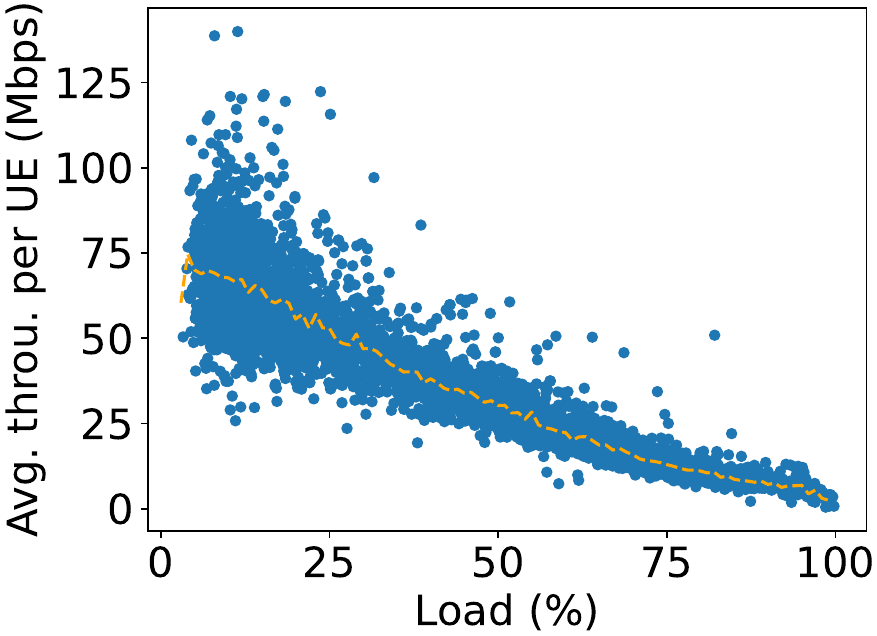}
        \caption{4G}
        \label{fig:scatter-thr}
    \end{subfigure}%
    \hspace{+1mm}
    \begin{subfigure}[t]{0.48\linewidth} 
        \centering
        \includegraphics[width=0.95\linewidth]{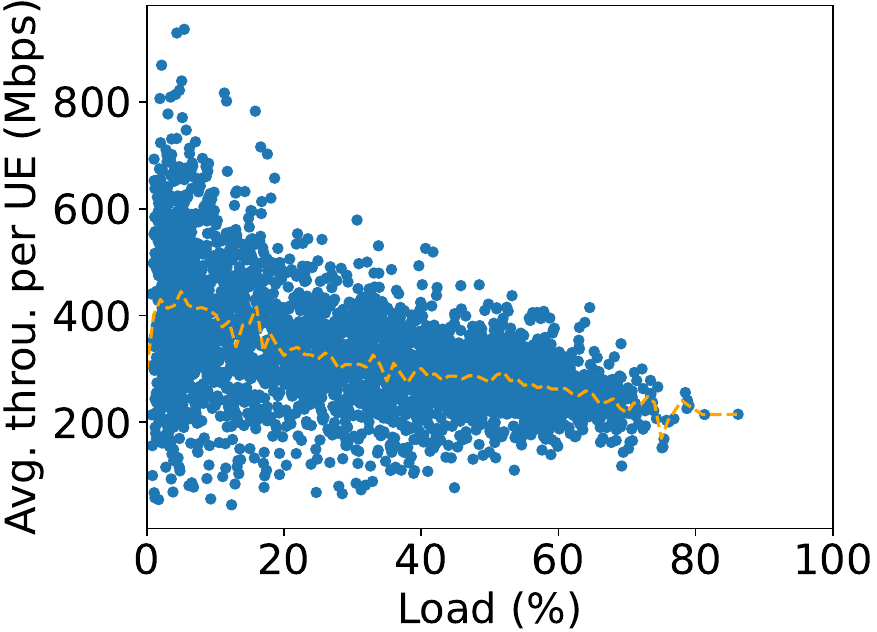}
        \caption{5G}
        \label{fig:scatter-thrb}
    \end{subfigure}
    \caption{Throughput vs load of a cell (scatter plot).}
    \label{fig:scatter-thr}
\end{figure}

{In \cite{ericsson}, the \textit{average throughput per UE} is defined as the data volume on a fifteen-minute interval divided by the time the scheduler actively assigns data to users. As shown in Figure \ref{fig:scatter-thr}, at low demand levels, when a small number of active UEs occupy a large portion of the available PRBs, the scheduler allocates data only intermittently. This results in an inflated measured throughput, as the data rate appears high during short bursts of activity. Conversely, as the cell load approaches 100\%, the scheduler continuously assigns data across all slots to meet demand, making the throughput KPI more indicative of the constant data rate that will be experienced by the UEs during the 15 minutes. }

To perform a QoS impact estimation we propose a methodology which is divided into two main phases: (i) Traffic offloading and balancing, and (ii) QoS estimation. First, we verify whether the 5G traffic demand can be offloaded to the 4G cells within the same sector and site, following the methodology detailed in previous subsection. If feasible, we apply a water-filling algorithm to balance the 5G load and UEs across all available 4G cells in the sector, assuming a uniform distribution of PRBs among UEs. This process produces an updated 4G load for each cell based on the portion of 5G traffic that has been offloaded.

Second, we utilize cell-specific look-up tables to estimate the average throughput per UE in each 4G cell based on the updated 4G loads after the traffic offloading process. The throughput estimation is based on the average throughput experienced at each load level, represented as the orange dashed line in Figure \ref{fig:scatter-thr}. 



To provide insights into the trade-off between switch-off opportunities and QoS, we define a QoS-aware switch-off strategy that, from the switch-off opportunities identified by the greedy strategy, considers only those where the UEs experience a specific average throughput. In our subsequent analysis, we consider an oracle-like implementation of this QoS-aware strategy that quantifies the resulting switch-off opportunities by having full visibility of the dataset. A practical implementation of this QoS-aware strategy is left for future work. 

We envision that an MNO could set the throughput requirement as part of a network-wide energy-saving policy. In this paper, those are derived from the typical requirements of popular video streaming applications like YouTube and Netflix. For example, YouTube\footnote{\url{https://support.google.com/youtube/answer/78358?hl=en}} recommends a minimum of 5 Mbps for FHD 1080p videos and 20 Mbps for UHD 4K, while Netflix\footnote{\url{https://help.netflix.com/en/node/306}} requires at least 15 Mbps for UHD 4K streaming. Applying these requirements for the example cell over the analyzed week we obtain the following switch-off times: (i) 53\%  when considering 0 Mbps or 5 Mbps (energy savings upper bound), (ii) 45.6\%  for 10 Mbps, (iv) 36.6\% of time for 15 Mbps, (v) 33.4\% of time for 20 Mbps, and (vi) 28.6\% of time for 25 Mbps. As expected, the time percentage meeting each throughput level decreases significantly as the required throughput is increased.

\section{Network-Wide Evaluation}\label{sec:eval}

{In this section, we apply the proposed framework to all the cells in the analyzed area: 848 4G cells and 195 5G cells at 3500 MHz. First, we present the results using the default  $\gamma = 100\%$ of 4G occupancy, which provides the upper bound for energy savings.  Then, we analyze the trade-off between energy savings and QoS. Finally, we discuss the effects of lowering $\gamma$ to increase the percentage of cells meeting specific QoS requirements during the switch-off time. This latter analysis offers valuable insights into the design of intelligent strategies that effectively balance energy savings and QoS.}

\subsection{Analysis of energy-saving opportunities}

{First, we characterize the energy-saving opportunities considering the greedy switch-off strategy with $\gamma = 100\%$ over the complete cluster of cells. Figure \ref{fig:opport-energy} uses a heat map to illustrate the percentage of time each cell could be deactivated during the analyzed week. On average, this corresponds to the 79\% of the week, resulting in 13.7 MWh of energy savings. However, as shown in the figure, there is a high variability in the switch-off time of individual cells, some of them reaching 100\% of the time while others not surpassing the 20\% of time. Concretely, 64 cells (i.e., approximately 33\% of the cells) could be switched off the entire week, representing energy savings of 5.7 MWh. Without considering these cells, the average network-wide switch-off time decreases to 68.5\%. 

The heat map shows that switch off times across the different cells follow no specific pattern. In general, the dark cells (low-opportunities cells) are found to be in social or cultural points of interest, where cell load tends to be higher during the whole analyzed period. On the other hand, residential areas show higher energy-saving opportunities, probably due to the high penetration of fibre in this area which leads to a lower utilization of the cellular network

\begin{figure}[t]
    
\centering
\vspace{0.04in}
    \includegraphics[width=.7\linewidth]{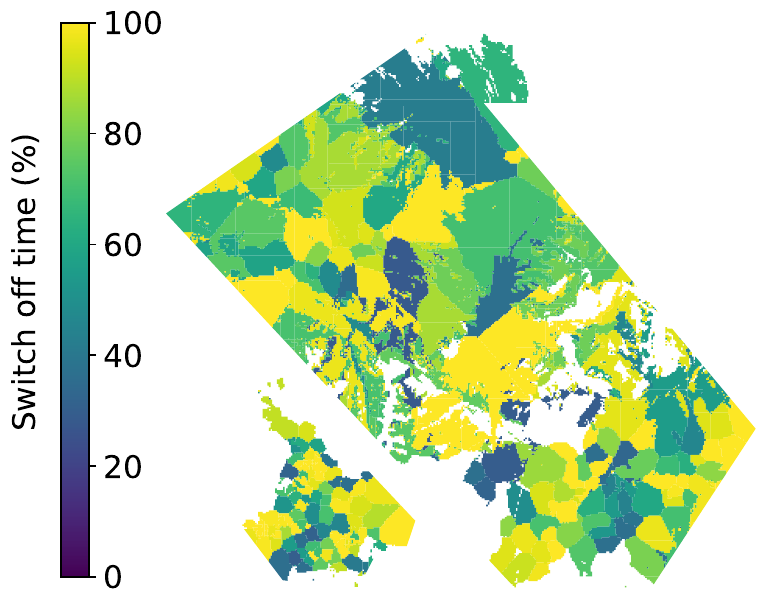}
    \caption{Calculated switch-off time (\% of the analyzed week).}
        \label{fig:opport-energy}

\end{figure}

\begin{figure}[t]
    \centering
    \begin{subfigure}[t]{0.48\columnwidth} 
   \centering
    \includegraphics[width=0.95\linewidth]{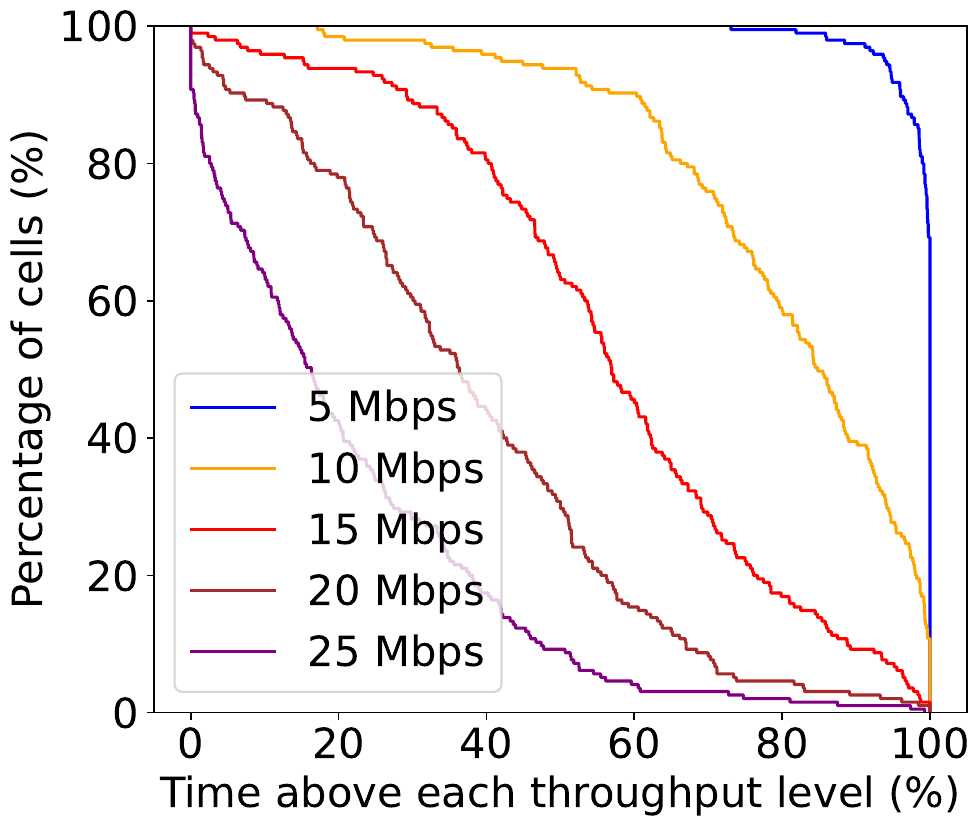}
    \caption{Percentage of cells meeting each defined QoS level over a certain time period ($\gamma$=100\%).}
    \label{fig:qos-cdfs}
        \end{subfigure}
    \hspace{+1mm}    \begin{subfigure}[t]{0.48\columnwidth} 
     \centering
    \includegraphics[width=0.95\linewidth]{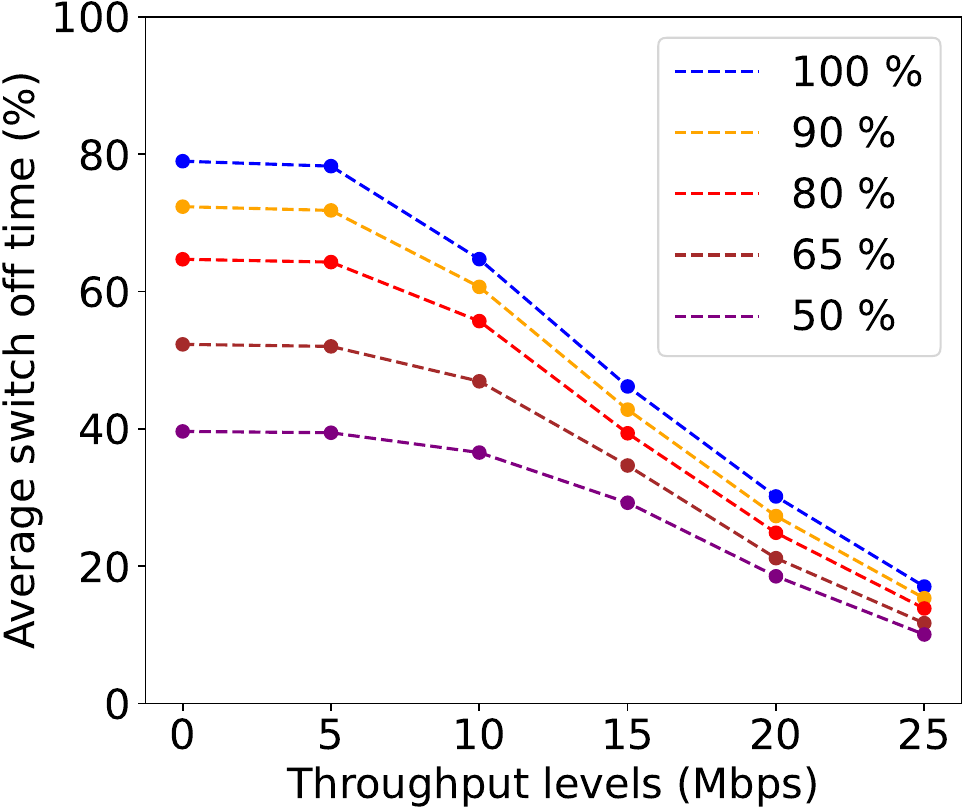}
    \caption{$\gamma$ impact on energy savings.}
    \label{fig:threshold-energy}
    \end{subfigure}
    \caption{Energy-saving and QoS trade-off according to $\gamma$.}
\end{figure}

\subsection{Trade-off between QoS and energy savings}

{Next, we evaluate the experienced QoS by the UEs when the greedy switch-off strategy. This is illustrated in the CCDF of Figure \ref{fig:qos-cdfs}, which depicts the percentage of cells meeting each defined QoS level over a certain time period. The values are normalized based on the time the cells are switched off, i.e., the 0 Mbps trend would be a vertical line at the 100\% level.
Note that increasing the required throughput level not only reduces the amount of cells able to provide it, but also the amount of time those can maintain it. For instance, while all the cells can provide 5 Mbps during almost 80\% of the switch-off time, the percentage of cells drops to less than 60\% and less than 20\% for 10 Mbps and 15 Mbps, respectively. Conversely, nearly 80\% of the cells can provide 5 Mbps during the entire switch-off period, but this drops to 40\% and 20\% of the time for 15 Mbps and 20 Mbps, respectively.} 

{A high-level analysis on the energy-saving and QoS trade-off can be extracted from the blue line of Figure \ref{fig:threshold-energy}, which represents the average switch-off time achieved when ensuring a specific throughput level during the entire switch-off time. While using a 5 Mbps limit achieves energy savings comparable to the greedy switch-off strategy, or 0 Mbps limit, (78.3\% versus 79\%), we observe a drastic reduction as the throughput levels increase, dropping to 17\% when ensuring 25 Mbps, i.e., 2.9 MWh of energy savings.
This highlights the inherent trade-off between energy savings and the QoS experienced by users, which should be considered by operators when implementing such energy-saving strategies.  }



\subsection{Study of $\gamma$ impact on QoS and energy-saving trade-off}

\begin{figure}[t]
    \centering
        \begin{subfigure}[t]{0.48\columnwidth} 
     \centering
    \includegraphics[width=0.95\linewidth]{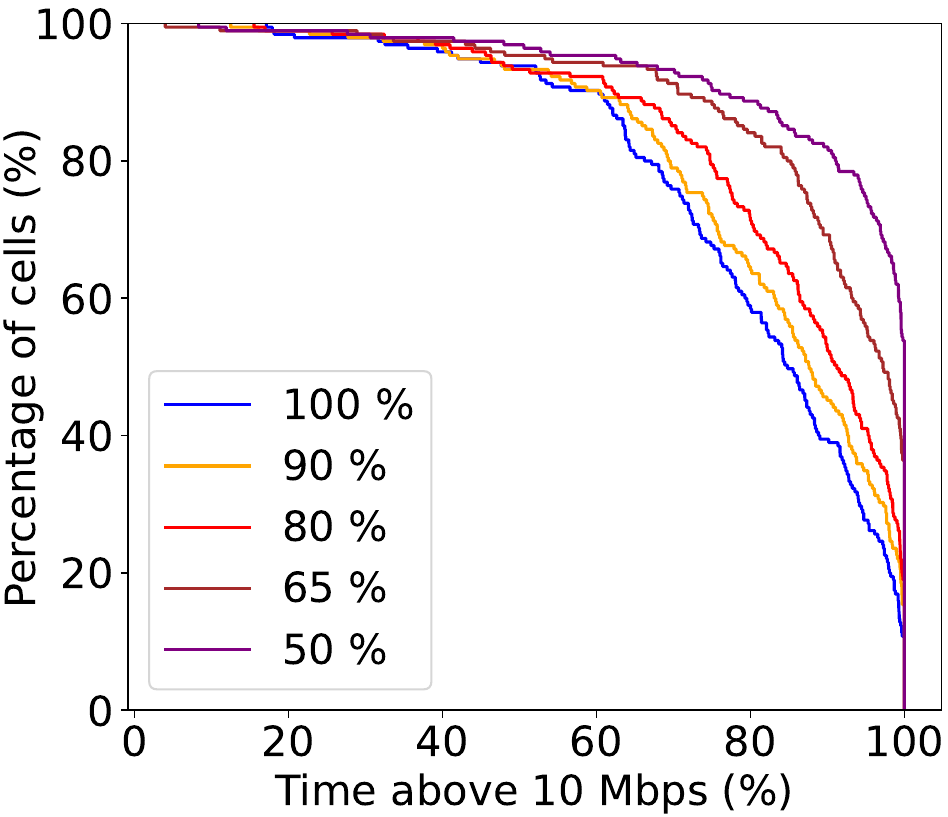}
    \caption{Threshold impact on time above 10 Mbps.}
    \label{fig:threshold-above-10}
    \end{subfigure}
        \begin{subfigure}[t]{0.49\columnwidth} %
   \centering
    \includegraphics[width=0.95\linewidth]{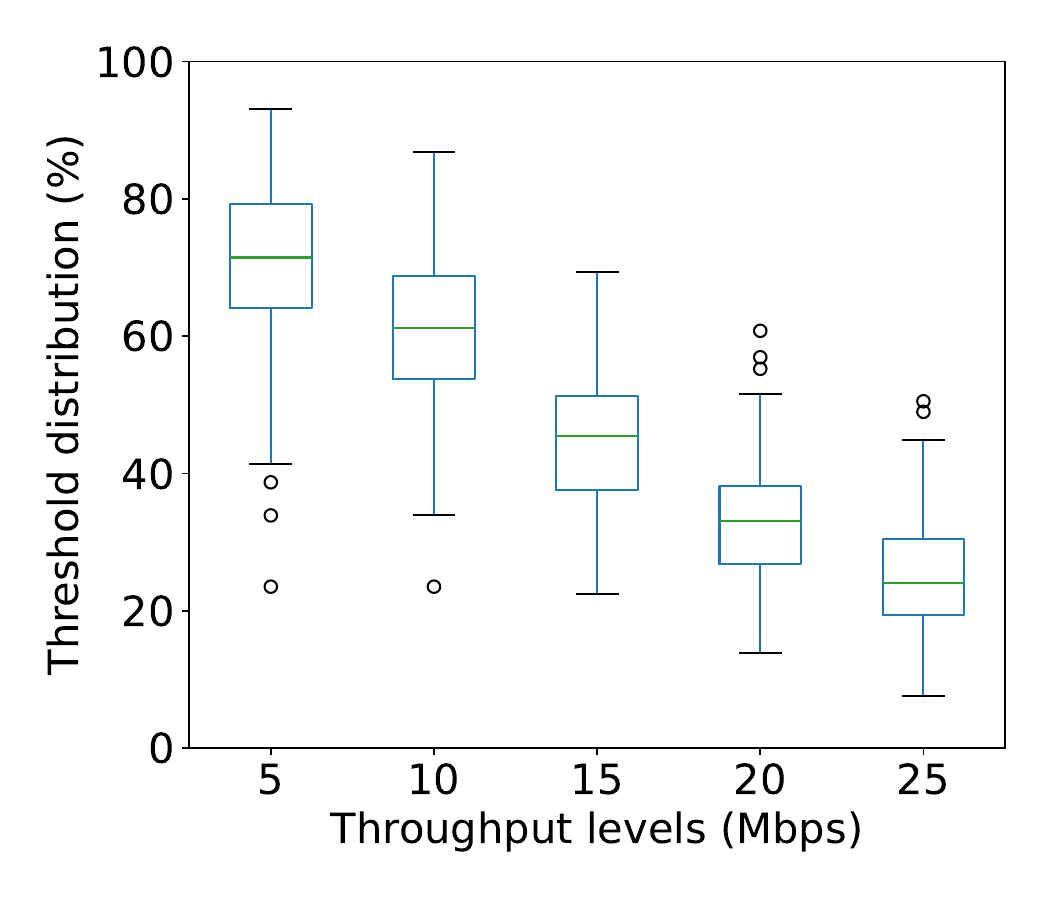}
    \caption{Cells $\gamma$ distribution for each throughput level.}
    \label{fig:boxplots}
        \end{subfigure}
    \caption{Impact of $\gamma$ on throughput levels.}
    \label{fig:5gvs4g}
\end{figure}

{In this section, we analyze the impact of lowering the $\gamma$ value. The objective is to determine whether reducing it can help to achieve specific QoS limits during the complete switch-off period.  
 {Lowering $\gamma$ forces the switch-on to occur earlier, and therefore its effect is twofold. On the one hand, Figure \ref{fig:threshold-above-10} shows that it effectively increases the percentage of cells able to maintain the 10 Mbps level for a greater portion of their switch-off time. On the other hand, Figure \ref{fig:threshold-energy} illustrates how the energy-saving opportunities get reduced for lower $\gamma$ values. We observe that the blue trend (or upper bound) in Figure \ref{fig:threshold-energy} is not achieved regardless of the QoS level, revealing that the $\gamma$ values required to meet specific QoS levels across different cells might differ from one to another.}

{To further explore this aspect, Figure \ref{fig:boxplots} illustrates the distribution across cells of the $\gamma$ values required to achieve the specified QoS requirements throughout the entire switch-off time period. For each throughput level, cases where the 5G cell remains off during the complete week are not considered (i.e., no $\gamma$ can be inferred). 
As expected, the average decreases as the QoS requirements increase, aligning with the intuitive notion that lowering $\gamma$ increases the probability of meeting certain QoS levels. However, a significant dispersion is observed within each level, emphasizing the need of setting cell-specific $\gamma$ values to maximize energy-saving opportunities while meeting the QoS constraints. This analysis is aligned with the results extracted from Figure \ref{fig:opport-energy} which showed a high variability between different cells, and states that a common $\gamma$-based strategy cannot be set to meet the target QoS while enhancing energy savings.

\section{Conclusions}

{In this work, we presented a framework to evaluate the trade-off between energy-saving opportunities and QoS in cell switch-off strategies for 5G NSA deployments based on traffic offloading from 5G to 4G cells. The study uses real-world data from a commercial MNO, enabling a realistic evaluation. A significant number of switch-off oppportunities were identified, providing an upper bound of 79\% of the week's time without considering QoS constraints. However, achieving these savings while maintaining a target service quality proved to be unfeasible. For instance, the average switch-off time decreased to 46\% and 30\% when 15 Mbps and 20 Mbps requirements were imposed, respectively. The analysis also highlighted the importance of considering site characteristics and context, such as cell location and available 4G carriers, thus discouraging the use of global fixed strategies.}

{To evaluate the achievable energy-QoS trade-off, in this paper we have considered an oracle-like implementation of the proposed cell switch-off strategies. Future research will focus on developing practically implementable strategies that effectively balance QoS and energy savings. To this end, AI/ML algorithms, such as predictive models for traffic forecasting and clustering techniques for cell-type aggregation, offer significant potential to optimize performance. Additionally, integrating digital twins to emulate the dataset scenario will enable the detailed consideration of link quality and mobility factors impacting the load distribution and the experienced QoS. Furthermore, O-RAN based traffic steering strategies could be considered to enable cross-sector approaches, increasing energy-saving opportunities.}



\bibliographystyle{IEEEtran}
\bibliography{references}

\end{document}